\documentclass[12pt]{article}
\usepackage[utf8]{inputenc}

\usepackage[margin=1in]{geometry}
\usepackage{amsmath, amssymb}
\usepackage{booktabs}
\usepackage{array}
\usepackage{tikz}
\usepackage{setspace}
\usepackage{hyperref}

\title{\textbf{Formation of Circular Directed Networks with Shared Link Costs}}
\author{Juan M. C. Larrosa\thanks{Department of Economics, Universidad Nacional del Sur; Instituto de Ciencias e Ingeniería de la Computación (ICIC). E-mail: \texttt{jlarrosa@uns.edu.ar}.}
\and Fernando A. Tohm\'e\thanks{Department of Economics, Universidad Nacional del Sur; Instituto de Matem\'atica de Bah\'ia Blanca (INMABB). E-mail: \texttt{ftohme@uns.edu.ar}.}}
\date{}

\begin{document}

\maketitle

\begin{abstract}
This paper develops a noncooperative model of directed network formation in which agents create links to access valuable information while sharing the costs generated along the paths through which information is obtained. Each agent is endowed with a positive amount of information and chooses, simultaneously, which other agents to contact. A directed link initiated by one agent allows her to access the information of the contacted agent and of the latter’s reachable network, but each link in the resulting information path entails a unit cost. Payoffs therefore depend on the total value of accessible information net of the accumulated connection costs required to obtain it. The paper characterizes the relationship between strategy profiles and directed graphs, defines accessibility, paths, components, and minimal connectedness, and studies the Nash architectures induced by individual best responses. The central result is that strict Nash equilibria must take the form of circular directed networks. Moreover, circular networks are exactly the Nash networks that use the minimum number of links while allowing every agent to access all available information. Although noncircular weak Nash networks may exist, they are structurally redundant and do not satisfy the same minimality property. The model also shows that strict Nash networks are both Pareto optimal and efficient in terms of aggregate welfare. Finally, the paper compares this framework with Bala and Goyal’s model, emphasizing that shared path costs and heterogeneous information values generate different equilibrium implications. The analysis supports the equivalence between strict stability and minimal connectivity in directed information networks.


\end{abstract}

\section*{Introduction}

Interaction among agents can be represented in several ways. One way of representing direct exchanges that has attracted considerable interest in recent years is through networks. Since it has a simple graphical representation, this analytical tool was first adopted in sociology and anthropology. For experts in those areas, it constitutes a graphical way of understanding the influence of agents' environments on individual behavior. Based on the study of real social networks, sociologists and anthropologists have accumulated extensive evidence that helps explain how human behavior is conditioned by the behavior of other agents.

In mathematical terms, a network is a graph, where nodes represent individual agents and arcs or links are interpreted as a ``utility good''---for example, information, personal prestige, and so forth---that is exchanged; see Wasserman and Faust (1994). Economic literature has recently introduced tools from game theory into this analytical framework. Rather than being interested only in descriptive aspects, some economic theorists have addressed the study of how networks are formed in the first place, and then their conditions of stability and efficiency; see Jackson and Wolinsky (1996), Bala and Goyal (2000), and Dutta and Jackson (2000). The game-theoretic approach to networks has two main strands: one based on cooperative games and the other on strategies.

The analysis based on cooperative games, as is usual in this approach, studies the problem of coalition formation among agents. The demanding assumption of utility transfer among agents is difficult to justify in many cases, in addition to being computationally costly; see Qin (1996), Dutta et al. (1998), and Slikker and van den Nouweland (2001).

In turn, the strategic or noncooperative approach only requires the definition of the strategies available to agents, as well as the characterization of the corresponding payoff functions. Given a certain protocol or rule of interaction, agents decide whether or not to connect to the network, evaluating the benefits of connection or disconnection with other agents. The rational decisions of the agents lead to Nash equilibria, which give rise to the networks in this analysis.

As mentioned, a network can be viewed as a graph. An important modeling decision is whether the graph is to be directed or undirected. This choice of primitives also has consequences for equilibrium results in noncooperative games of network formation. Undirected graphs are useful for representing situations in which the direction of flows of utility goods is less important or irrelevant; see Dutta and Mutuswami (1997). Directed graphs, on the other hand, reflect the importance of distinguishing which agent initiated the connection as well as the direction taken by the flow of information. A common convention is to draw directed links with arrows pointing toward the agent who decided to initiate the connection; see Bala and Goyal (2000) and Dutta and Jackson (2001).

In this paper, networks are designed as graphs with directed flows. We call the utility good that circulates through networks ``information,'' in a rather generic use of the term. Each agent is assigned some amount of information, but has a payoff function that depends positively on the amount of information to which she has access. By establishing links with others, the agent can acquire information, but she has to share the payment of the cost of information. This is a way of representing the frequent fact that indirectly obtained information nevertheless requires a certain amount of cooperation with the source, in order to provide incentives for the source to continue supplying it.

The shared-cost approach applied in this paper assumes that each agent pays a small fee for each link in the path that allows her to reach the desired information. The problem is to determine which structure can emerge as a strategic equilibrium among the agents and whether, in addition, it is optimal. It is shown here that strict Nash equilibria, or Nash equilibria with the minimum number of links, give rise to a circular network, which is stable and optimal.

The problem of network formation studied here can be observed in different situations. For instance, consider the following scenario: suppose that Internet users are charged a small amount for each link through which a website is visited. If one of them follows a link, she has to pay for the new connection but obtains access to more information. If she then follows a link on that site, she accesses the new site by again paying the fee, but obtaining access to more information. The question is: what is the most efficient way to navigate through a series of sites under this cost structure?

Closer to this framework, one might ask what kind of architecture for a local area network (LAN) increases the speed of flow while reducing losses. This is in fact analogous to our generic problem: a particular computer in the LAN might need to use the resources of another computer in the network. There should be an efficient protocol for choosing which machine to connect to. At the same time, a small ``fee'' must be paid---for example, in terms of processing time---in order to reach the machine that provides the largest amount of resources. Since this is true for all PCs in the network, the strategic outcome must allow all of them to reach the largest amount of available information while paying as little as possible. As in our result, although for technological reasons, the final outcome could be a circular network, as in the case of IBM's Token Ring architecture; see Tanenbaum (1989).

Such a situation appears in various social organizations, for example in multidisciplinary evaluation committees. These are made up of experts in different fields. Each one must rely on others to obtain information about areas in which she is not an expert. In this case, the circular network minimizes the number of questions while at the same time maximizing the information available to everyone.

The remainder of the paper is organized as follows. Section 2 presents the model. Section 3 determines the equilibrium architecture, also showing how the equilibria satisfy certain stability and optimality criteria. Section 4 discusses the analogies and differences with the original work of Bala and Goyal. Finally, Section 5 offers a brief evaluation of the results reported here.

\section*{The Model}

Let $N=\{1,\ldots,n\}$ be a set of agents. To avoid trivial results, assume that $n\geq 3$. If $i$ and $j$ are two typical members of $N$, a link between them, without intermediaries, originated in $i$ and ending in $j$, will be represented as $ij$. The interpretation of $ij$ is that $i$ establishes contact with $j$, allowing $i$ to access $j$'s information as well as her network of contacts. Each agent $i\in N$ has some information of her own, $I_i\in \mathbb{Z}_{+}$, that is, represented as a positive integer. As mentioned, $i$ can access more information by forming links with other agents. Network formation is costly in time, resources, and effort, but for simplicity we shall assume that the link $ij$ has a cost of 1, measured in units of information utility. By convention, it is assumed that the information of each agent is sufficiently valuable for it to be worthwhile to establish a link with her, that is, $I_i>1$.

Agents will try to maximize the utility of the available information while minimizing the cost of connection with other agents. To achieve this, they will choose one strategy from a set of strategies. Each strategy for $i\in N$ is an $(n-1)$-dimensional vector
\[
g_i=(g_{i,1},\ldots,g_{i,i-1},g_{i,i+1},\ldots,g_{i,n}),
\]
where each $g_{i,j}$, for $j\neq i$, takes the value 0 or 1. This is interpreted as follows: $i$ establishes a direct link with $j$ if $g_{i,j}=1$, whereas if $g_{i,j}=0$ such a link does not exist. The set of all strategies is denoted by $G_i$. The analysis is restricted to pure strategies, which implies that $|G_i|=2^{n-1}$. Finally,
\[
G=G_1\times\cdots\times G_n
\]
denotes the set of strategy profiles in the interaction among the agents in $N$.

The existence of a direct link $ij$ indicates asymmetric communication between $i$ and $j$. That is, $g_{i,j}=1$ indicates that $i$ has established communication with $j$, which allows $i$ to access $j$'s information, but not vice versa. Symmetry between $i$ and $j$ is restored if $g_{j,i}=1$. Structures with this characteristic are called directed-flow networks.

In these networks, the strategy profile can be represented as a directed graph
\[
g=(g_1,\ldots,g_n)
\]
on $N$. That is, in the directed graph the elements of $N$ are the nodes, while each link established as $g_{i,j}=1$ is represented by an arrow beginning at $j$ and directed toward $i$. This represents the idea that when $i$ establishes a link with $j$, information flows from $j$ to $i$. Thus, arrows are always oriented toward the agent who establishes the link. It follows immediately that:

\textbf{Proposition 1.} There is a bijective relationship between directed graphs among $n$ nodes and strategy profiles in $G$.

\textit{Proof.} A directed graph with $n$ nodes is such that, for each node $i$, there is at most one incoming arrow from each $j\neq i$, and none from itself. Therefore, for each $j$, define $g_{i,j}$ equal to 1 if there is an incoming arrow from $j$, and 0 otherwise. This defines
\[
g_i=(g_{i,1},\ldots,g_{i,i-1},g_{i,i+1},\ldots,g_{i,n})
\]
for each $i\in N$, and a
\[
g=(g_1,\ldots,g_i,\ldots,g_n)\in G.
\]
Likewise, given a $g$, a directed graph can be obtained by adding an arrow from $j$ to $i$ if $g_{i,j}=1$. Since $g_{i,i}$ is not defined, the graph has no self-loops, and since $g_{i,j}$ has only two possible values, there is either one link between them or none. \hfill $\square$

\textbf{Example 1.} Given a group of four agents,
\[
N=\{a,b,c,d\},
\]
a joint strategy
\[
g=(g_a,g_b,g_c,g_d)
\]
can be represented as a strategy profile, as in Table 1.

Each row is the strategy chosen by one of the agents. The columns correspond to the agents. An entry 1 in row $i$ and column $j$ means that the strategy of agent $i$ prescribes establishing a link with agent $j$. Entries on the main diagonal are marked with crosses, since agents cannot establish links with themselves. Figure 1 shows the directed graph corresponding to $g$.

\begin{table}[h!]
\centering
\caption{Strategy profile}
\begin{tabular}{c|cccc}
\toprule
Strategy & $a$ & $b$ & $c$ & $d$ \\
\midrule
$g_a$ & X & 1 & 0 & 0 \\
$g_b$ & 0 & X & 1 & 0 \\
$g_c$ & 0 & 0 & X & 1 \\
$g_d$ & 0 & 0 & 0 & X \\
\bottomrule
\end{tabular}
\end{table}

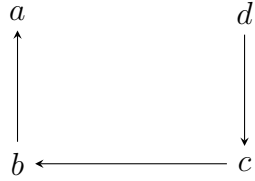
\begin{figure}[h!]
\centering
\begin{tikzpicture}[>=stealth, node distance=2.5cm]
\node (a) at (0,2) {$a$};
\node (b) at (0,0) {$b$};
\node (c) at (3,0) {$c$};
\node (d) at (3,2) {$d$};
\draw[->] (b) -- (a);
\draw[->] (c) -- (b);
\draw[->] (d) -- (c);
\end{tikzpicture}
\caption{Network formed by the strategy profile}
\end{figure}

Define
\[
N^{g_i}=\{k\in N\mid g_{i,k}=1\}
\]
as the set of agents with whom $i$ establishes a direct link according to her strategy profile $g_i$. There is a path from $j$ to $i$ according to $g\in G$ if there is a sequence of different agents, to avoid cycles,
\[
j_0,\ldots,j_m,
\]
with $i=j_0$ and $j=j_m$, such that
\[
g_{j_0,j_1}=\cdots=g_{j_{m-1},j_m}=1.
\]
Given a joint strategy $g$, we have
\[
j_1\in N^{g_{j_0}},\quad j_2\in N^{g_{j_1}},\quad \ldots,\quad j_m\in N^{g_{j_{m-1}}}.
\]
A path from $j=j_m$ to $i=j_0$, denoted as $j\to i$, has a length equal to the cardinality of the sequence
\[
j_1,j_2,\ldots,j_{m-1},j_m,
\]
that is, $m$, which indicates the number of intermediate links between $j$ and $i$. Notice that a directed link is a path of length 1.

\textbf{Example 1, reformulated.} Given the strategy
\[
g=(g_a,g_b,g_c,g_d),
\]
with
\[
N^{g_a}=\{b\},\quad N^{g_b}=\{c\},\quad N^{g_c}=\{d\},
\]
while
\[
N^{g_d}=\varnothing.
\]
This sequence establishes a path from $d$ to $a$ of length 3.

The set of agents accessed, directly or otherwise, by $i$ is denoted as
\[
N^{i;g}=\{k\in N\mid k\to i\}\cup\{i\}.
\]
The agent $i$ is included in $N^{i;g}$ to indicate that $i$ knows her own valuation, despite the previously mentioned fact that $i$ does not establish a direct link with herself. Let
\[
\mu_i:G\to \{0,\ldots,n(n-1)\}
\]
be the number of links in all the paths that end in $i$, originated by agents in $N^{i;g}$ under any joint strategy:
\[
\mu_i(g)=\left|\left\{(j,k)\in N\times N:
g_{j,k}=1,\ \text{and there exists } l\in N^{i;g} \text{ with } l\to i
\text{ and } j,k\in l\to i
\right\}\right|.
\]
Notice that there may be more than one path from $j$ to $i$.

\textbf{Example 2.} Suppose now that
\[
N=\{1,2,3,4,5\}
\]
and the strategy
\[
g=(g_1,g_2,g_3,g_4,g_5)
\]
is given by Table 2.

\begin{table}[h!]
\centering
\caption{Strategy profile}
\begin{tabular}{c|ccccc}
\toprule
Strategy & 1 & 2 & 3 & 4 & 5 \\
\midrule
$g_1$ & X & 1 & 0 & 0 & 1 \\
$g_2$ & 0 & X & 1 & 0 & 0 \\
$g_3$ & 0 & 0 & X & 1 & 1 \\
$g_4$ & 0 & 0 & 0 & X & 0 \\
$g_5$ & 0 & 0 & 0 & 0 & X \\
\bottomrule
\end{tabular}
\end{table}

\begin{figure}[h!]
\centering
\begin{tikzpicture}[>=stealth]
\node (one) at (0,2) {1};
\node (two) at (0,0) {2};
\node (three) at (3,0) {3};
\node (four) at (6,0) {4};
\node (five) at (3,2) {5};
\draw[->] (two) -- (one);
\draw[->] (three) -- (two);
\draw[->] (four) -- (three);
\draw[->] (five) -- (three);
\draw[->] (five) -- (one);
\end{tikzpicture}
\caption{Network formed by the strategy profile}
\end{figure}
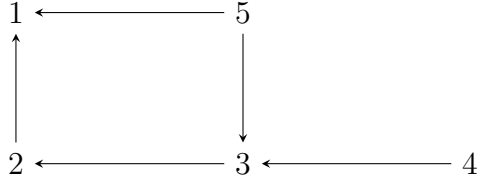

Figure 2 shows the corresponding network. Here
\[
N^{1;g}=\{1,2,3,4,5\},\quad
N^{2;g}=\{2,3,4,5\},
\]
\[
N^{3;g}=\{3,4,5\},\quad
N^{4;g}=\{4\},\quad
N^{5;g}=\{5\}.
\]
That is, under $g$, agent 1 accesses the information of all agents, whereas agents 4 and 5 access only their own information. The numbers of links required to obtain the information are
\[
\mu_1(g)=5,\quad \mu_2(g)=3,\quad \mu_3(g)=2,
\]
while
\[
\mu_4(g)=\mu_5(g)=0.
\]

To turn this scheme into a game, the agents' payoffs are defined. We shall assume that
\[
\Pi_i:G\to\mathbb{R},
\]
the payoff function for agent $i$, is:
\begin{equation}
\Pi_i(g)=\sum_{j\in N^{i;g}} I_j-\mu_i(g).
\tag{1}
\end{equation}

That is, the payoffs of $i$ are the sum of all the information to which she has access, minus the cost of the paths reaching her, established according to $g$, recalling that each link has unit cost. The intuition here is that $i$ obtains a payoff from accessing more information, but at the same time she must pay a charge or fee for each of the links in the paths to the sources of information.

\textbf{Example 2, first reformulation.} Suppose that the information obtained by the agents is:
\[
I_1=2,\quad I_2=2,\quad I_3=4,\quad I_4=3,\quad I_5=3.
\]
Then, under the strategy $g$:
\[
\Pi_1(g)=I_1+\cdots+I_5-\mu_1(g)=2+2+4+3+3-5=9,
\]
\[
\Pi_2(g)=I_2+\cdots+I_5-\mu_2(g)=2+4+3+3-3=9,
\]
\[
\Pi_3(g)=I_3+\cdots+I_5-\mu_3(g)=4+3+3-2=8,
\]
\[
\Pi_4(g)=I_4-\mu_4(g)=3-0=3,
\]
\[
\Pi_5(g)=I_5-\mu_5(g)=3-0=3.
\]

We can note that, for example, if $g_{1,5}=0$, agent 1 could improve her payoff, obtaining 10 rather than 9, since she would still have access to $I_5$ but using one less link.

For each $g\in G$, agent $i$ obtains a structure $N^{i;g}$, and her payoff depends critically on the type of graph corresponding to $N^{i;g}$, as summarized in the following proposition.

\textbf{Proposition 2.} Given two joint strategies $g$ and $g'$, $\Pi_i(g)\geq \Pi_i(g')$ if and only if the corresponding graphs $N^{i;g}$ and $N^{i;g'}$ are such that:
\[
\sum_{j\in N^{i;g}} I_j-\sum_{j\in N^{i;g'}} I_j
\geq
\mu_i(g)-\mu_i(g').
\]

\textit{Proof.} Trivial. \hfill $\square$

This result helps to understand the presumption that the objective of a rational agent is to obtain as much information as possible while crossing the smallest possible number of links. There are two cases of particular interest:
\[
\sum_{j\in N^{i;g}} I_j=\sum_{j\in N^{i;g'}} I_j
\quad \text{and}\quad
\mu_i(g)\leq \mu_i(g'),
\]
\[
\sum_{j\in N^{i;g}} I_j\geq \sum_{j\in N^{i;g'}} I_j
\quad \text{and}\quad
\mu_i(g)=\mu_i(g').
\]

The first condition shows that $\Pi_i(g)\geq \Pi_i(g')$ if the information obtained through $g$ is the same as that obtained through $g'$, but the number of required links is smaller in $g$ than in $g'$. The second case shows that $\Pi_i(g)\geq \Pi_i(g')$ if the number of links required to reach the information is the same in $g$ as in $g'$, but the amount of information obtained in $g$ is greater than that reached in $g'$.

\section*{Equilibrium and Optimality}

Given a network $g\in G$, which according to Proposition 1 corresponds to a joint strategy $g$ with its corresponding directed graph, let $g_{-i}$ be the directed graph obtained when all direct links of agent $i$ are removed. Then $g$ can be written as
\[
g=(g_i,g_{-i}),
\]
meaning that $g$ is formed by the union of the links in $g_i$ and those in $g_{-i}$. A strategy $g_i$ is said to be a best response of agent $i$ to $g_{-i}$ if
\begin{equation}
\Pi_i(g_i,g_{-i})\geq \Pi_i(g_i',g_{-i})
\tag{2}
\end{equation}
for every $g_i'\in G_i$.

\textbf{Example 3.} Consider again the case
\[
N=\{1,2,3,4,5\},
\]
where
\[
I_1=2,\quad I_2=2,\quad I_3=4,\quad I_4=3,\quad I_5=3.
\]
Let $g_{-1}$ be described by Table 3. Also see Figure 3 for the situation faced by agent 1.

\begin{table}[h!]
\centering
\caption{Strategy profile}
\begin{tabular}{c|ccccc}
\toprule
Strategy & 1 & 2 & 3 & 4 & 5 \\
\midrule
$g_2$ & 0 & X & 1 & 0 & 1 \\
$g_3$ & 0 & 0 & X & 1 & 0 \\
$g_4$ & 0 & 0 & 0 & X & 0 \\
$g_5$ & 0 & 0 & 0 & 0 & X \\
\bottomrule
\end{tabular}
\end{table}

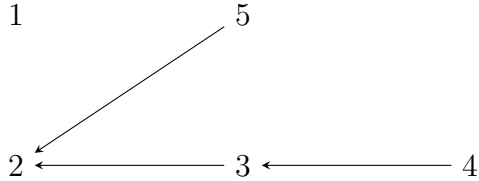
\begin{figure}[h!]
\centering
\begin{tikzpicture}[>=stealth]
\node (one) at (0,2) {1};
\node (two) at (0,0) {2};
\node (three) at (3,0) {3};
\node (four) at (6,0) {4};
\node (five) at (3,2) {5};
\draw[->] (three) -- (two);
\draw[->] (four) -- (three);
\draw[->] (five) -- (two);
\end{tikzpicture}
\caption{Network formed by the strategy profile}
\end{figure}

Agent 1 has to decide with whom to establish a connection. One possibility is to remain isolated, but that would give her only a payoff of 2. Alternatively, she could connect to as many agents as she wishes. But some connections might be redundant in terms of informational gains. Such redundancy, in turn, would imply a higher cost for the same information. Thus, for example, connecting to 3 and 4 would ensure that agent 1 has access to the information held by them. The number of required links would be 3. The payoff would therefore be
\[
2+4+3-3=6.
\]
She could instead connect only to 3, since she would still receive the information of 3 and 4 but would require only 2 links; that is, her payoff would be
\[
2+4+3-2=7.
\]
It can be deduced that the best response for 1 would be to connect only to the agent with the highest payoff under $g_{-1}$. This is agent 2, who has a payoff of
\[
2+4+3+3-3=9.
\]
Therefore, 1 will reach the information of 2, 3, 4, and 5, requiring 4 links. Thus, her payoff would be 10. Figure 4 shows the resulting network.

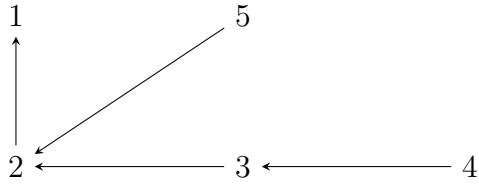
\begin{figure}[h!]
\centering
\begin{tikzpicture}[>=stealth]
\node (one) at (0,2) {1};
\node (two) at (0,0) {2};
\node (three) at (3,0) {3};
\node (four) at (6,0) {4};
\node (five) at (3,2) {5};
\draw[->] (two) -- (one);
\draw[->] (three) -- (two);
\draw[->] (four) -- (three);
\draw[->] (five) -- (two);
\end{tikzpicture}
\caption{Final network formed by agent 1}
\end{figure}

The set of best responses to $g_{-i}$ is $BR_i(g_{-i})$. A network
\[
g=(g_1,\ldots,g_n)
\]
is said to be a Nash network if, for each $i$,
\[
g_i\in BR_i(g_{-i}),
\]
that is, if $g$, as a joint strategy, is a Nash equilibrium. To determine the structure of Nash networks, several definitions are introduced that will make it possible to describe additional properties of networks.

Given a network $g$, a set $C\subset N$ is called a component of $g$ if, for every pair of agents $i$ and $j$ in $C$, with $i\neq j$, it holds that $j\in N^{i;g}$, and there is no $C'$, with $C\subset C'$, for which this is true. A component $C$ is said to be minimal if $C$ ceases to be a component once $g_{i,j}=1$ between two agents $i$ and $j$ in $C$ is interrupted, that is, if $g_{i,j}=0$.

\textbf{Example 4.} Given
\[
N=\{1,2,3,4\},
\]
consider the following network, represented in Table 4 and Figure 5.

\begin{table}[h!]
\centering
\caption{Strategy profile}
\begin{tabular}{c|cccc}
\toprule
Strategy & 1 & 2 & 3 & 4 \\
\midrule
$g_1$ & X & 1 & 0 & 0 \\
$g_2$ & 0 & X & 1 & 0 \\
$g_3$ & 0 & 0 & X & 1 \\
$g_4$ & 0 & 1 & 0 & X \\
\bottomrule
\end{tabular}
\end{table}

\begin{figure}[h!]
\centering
\begin{tikzpicture}[>=stealth]
\node (one) at (0,2) {1};
\node (two) at (0,0) {2};
\node (three) at (3,0) {3};
\node (four) at (3,2) {4};
\draw[->] (two) -- (one);
\draw[->] (three) -- (two);
\draw[->] (four) -- (three);
\draw[->] (two) -- (four);
\end{tikzpicture}
\caption{Network formed by the strategy profile}
\end{figure}
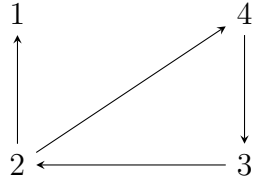

Clearly,
\[
C=\{2,3,4\}
\]
is a component, since
\[
N^{2;g}=N^{3;g}=N^{4;g}=\{2,3,4\}.
\]
If $N=C\cup\{1\}$ is considered, $N$ is not a component, since 1 does not belong to $N^{2;g}$, $N^{3;g}$, or $N^{4;g}$. On the other hand, $C$ is minimal, since if any of the links 23, 34, or 42 is interrupted, one of the agents ceases to be reachable for at least one agent in $C$. Thus, for example, if 23 is cut, in the new network $g'$,
\[
N^{2;g'}=\{2\}.
\]

A network is said to be connected if it has a single component. If that single component is minimal, $g$ is said to be minimally connected. A network that is not connected is said to be disconnected. A particular instance of minimally connected networks is the circular network, in which agents can be labeled, by means of a function $\ell:N\to N$, as
\[
\{\ell(1),\ldots,\ell(n)\}
\]
and
\[
g_{\ell(1),\ell(2)}=
g_{\ell(2),\ell(3)}=
\cdots =
g_{\ell(n-1),\ell(n)}=
g_{\ell(n),\ell(1)}=1,
\]
with no other links.

With all these elements, the following result can be established. All results in this section correspond to the game $(N,G,\Pi)$, where
\[
\Pi=\Pi_1\times\cdots\times\Pi_n.
\]

\textbf{Lemma 1.} If $g^*$ is a strict Nash network, then it is circular.

\textit{Proof.} Consider
\[
\Pi_i:G\to \mathbb{Z}
\]
for each $i\in N$, and a strict Nash equilibrium $g^*\in G$. Then, for each $i$ and each $g_i\in G_i$,
\begin{equation}
\Pi_i(g_i^*,g_{-i}^*)>\Pi_i(g_i,g_{-i}^*).
\tag{3}
\end{equation}

Consider the payoff associated with a circular network. If $g^*$ defines such a network, then, by Proposition 2, it must hold for each $i$ that:
\begin{equation}
\Pi_i(g^*)=\sum_{j\in N} I_j-(n-1).
\tag{4}
\end{equation}

In words: the maximum amount of information that can be reached in a circular network is the sum of the information held by all agents, while the number of links that make this information available to any one of them is $n-1$. Notice that a structure in which there is only one path between any pair of agents has only $n$ links.

Suppose, by contradiction, that $g^*$ is not circular. This means that for at least one agent $i$,
\begin{equation}
\Pi_i(g^*)\neq \sum_{j\in N} I_j-(n-1).
\tag{5}
\end{equation}

First consider the case in which:
\begin{equation}
\Pi_i(g^*)>\sum_{j\in N} I_j-(n-1).
\tag{6}
\end{equation}
Since $\sum_{j\in N} I_j$ cannot be improved, the only possibility is that the number of links is smaller, that is,
\[
\Pi_i(g^*)=\sum_{j\in N} I_j-k,
\]
where $k<n-1$. But a contradiction appears from the fact that $k\geq n-1$, since otherwise $i$ would not be able to access at least one agent $j$ and therefore could not obtain the benefit from her information $I_j$.

Now consider the case in which:
\begin{equation}
\Pi_i(g^*)<\sum_{j\in N} I_j-(n-1).
\tag{7}
\end{equation}
This can occur if $i$ does not have access to at least one agent, say $j$, or if the number of links in the paths to the information acquired by $i$ is greater than $n-1$. Consider the first case, that is, that there exists a $j$ who is not accessed by $i$. Then $i$ can select a strategy $g_i\in G_i$ such that $g_{i,j}=1$. The number of links then increases by 1, while the accessed information increases by $I_j>1$. Thus,
\[
\Pi_i(g_i,g_{-i}^*)>\Pi_i(g^*).
\]
This is absurd, since $g^*$ is a strict Nash equilibrium.

On the other hand, if the number of links in the path that provides information to $i$ is greater than $n-1$, $i$ receives the information of at least one agent $j$, $I_j$, in a redundant way. This implies that there is an agent $k$, which may be $i$ itself, such that $k$ receives information from $j$ both through a direct link, $g_{k,j}=1$, and through a link to another agent, say $l$. Then $k$ can switch to an alternative strategy $\bar g_k\in G_k$, identical to $g_k^*$ except for $\bar g_{k,j}=0$. This implies that the information accessed by $k$ is the same as under $g^*$, while the number of links is reduced by 1. Hence,
\[
\Pi_k(\bar g_k,g_{-k}^*)>\Pi_k(g^*).
\]
This is absurd, since $g^*$ is a Nash equilibrium. \hfill $\square$

Notice that not every Nash network is circular.

\textbf{Example 5.} Let
\[
N=\{1,2,3\}
\]
with
\[
I_i=2
\]
for $i=1,2,3$. Let $g^*$ be represented by Table 5.

\begin{table}[h!]
\centering
\caption{Strategy profile}
\begin{tabular}{c|ccc}
\toprule
Strategy & 1 & 2 & 3 \\
\midrule
$g_1^*$ & X & 1 & 1 \\
$g_2^*$ & 1 & X & 0 \\
$g_3^*$ & 1 & 0 & X \\
\bottomrule
\end{tabular}
\end{table}

Of course, $g^*$ does not define a circular network; this type is called a star network. To verify that $g^*$ is a weak Nash equilibrium, consider, for example, the best responses of 2 to $g_{-2}^*$; the analysis for 1 and 3 is analogous. Apart from $g_2^*$, there are three other possibilities:
\[
g_2^1=(0,X,0),\quad
g_2^2=(1,X,1),\quad
g_2^3=(0,X,1).
\]
Then, while
\[
\Pi_2(g^*)=2+2+2-2=4,
\]
we have
\[
\Pi_2(g_2^1,g_{-2}^*)=2,
\]
\[
\Pi_2(g_2^2,g_{-2}^*)=2+2-1=3,
\]
and
\[
\Pi_2(g_2^3,g_{-2}^*)=2+2+2-2=4.
\]

This example shows that there are noncircular Nash networks that can yield the same payoff as circular ones. But notice that while individuals obtain the same payoff, the global structure differs. In fact, the following holds.

\textbf{Proposition 3.} $g^*$ is a circular network if and only if it is a Nash network with the minimum number of links.

\textit{Proof.} If $g^*$ is a circular network, then for each $i$ the payoff is given by (4). Suppose that it is not a Nash network. That is, for at least one agent $i$, there exists a deviation $g_i'\in G_i$ such that
\begin{equation}
\Pi_i(g_i',g_{-i}^*)>\sum_{j\in N} I_j-(n-1).
\tag{8}
\end{equation}
However, the only way to reach this result is by reducing the number of links in the paths that carry information from the other agents to $i$. Since $g^*$ is circular, there is only one agent $j$ such that $g_{i,j}^*=1$. There are three possible deviations for $g_i'$.

First, for every $j\neq i$, let $g_{i,j}'=0$. Then $i$ reduces the number of links by $n-1$ and accesses only her own information, losing the information of all the other $n-1$ agents. Since
\[
\sum_{j\neq i} I_j>n-1,
\]
then
\[
\Pi_i(g_i',g_{-i}^*)<\sum_{j\in N}I_j-(n-1).
\]
This is a contradiction.

Second, for a given $k$, let $g_{i,k}'=1$ while it is zero for every other agent. Then $i$ cuts the entire path $k\to i$ that passes through $j$. If the length of this path is $m$, then $m-1$ agents are no longer accessed. Since the information lost is greater than the reduction in links, the payoff cannot improve. This again contradicts (8).

Third, for more than one $k$, let $g_{i,k}'=1$. Then, even if the number of accessed agents remains the same, the number of links increases. Thus,
\[
\Pi_i(g_i',g_{-i}^*)\leq \Pi_i(g^*),
\]
which is again a contradiction.

Now suppose that $g^*$ is a Nash equilibrium with the minimum number of links. Then it constitutes a single component that includes all agents in $N$; otherwise, the information of agents who are not accessed would be lost for at least one other agent, while the reduction in link costs would not be sufficient to compensate for that loss. Recall that each $I_i$ is greater than the cost of one link. As shown above, the minimum number of links that allows all agents to be connected is $n$.

To show that $g^*$ is circular, suppose that it is not. Then, for every labeling function $\ell:N\to N$, at least one of
\[
g_{\ell(1),\ell(2)},\,
g_{\ell(2),\ell(3)},\,
\ldots,\,
g_{\ell(n-1),\ell(n)},\,
g_{\ell(n),\ell(1)}
\]
has value 0, or else there exists another link. The latter possibility must be discarded, since $g^*$ has only $n$ links. Therefore, it must not be possible to connect all agents in $N$ in such a way that each agent is connected with only one agent.

But since $g^*$ must include all agents and connect them with $n$ links, it is possible to choose one of the agents in the structure, for example $i$, and assign to it the label $\ell(i)=1$. Agent $i$ is connected to only one agent $j$, since if $i$ were connected to two different agents, only $n-2$ links would remain to connect the other $n-1$ agents. In that case, at least one of the agents would not have a direct link directed toward her and would therefore obtain a payoff lower than the maximum. Accordingly, label the agent connected to $i$ as $j$, so that $\ell(j)=2$. Consider the only agent to whom $j$ is connected, say $k$. Label $k$ as $\ell(k)=3$. Proceed in the same way until the agent accessed by the path of connections, say $r$, is such that $\ell(r)=n$. Then, up to that point, $n-1$ links will have been accessed. It remains to establish to whom $r$ will connect. It cannot be any of the agents denoted as $2,\ldots,n-1$, since each of them has only one connection, toward the preceding agent. On the other hand, $r$ cannot connect to herself, since her payoff would be only $I_r$. Therefore, she must connect with $i$, who has label 1. Hence, there exists a labeling function $\ell$ such that
\[
g_{\ell(1),\ell(2)}=
g_{\ell(2),\ell(3)}=
\cdots =
g_{\ell(n-1),\ell(n)}=
g_{\ell(n),\ell(1)}=1.
\]
This contradicts the assumption that the network is not circular. Therefore, $g^*$ is circular.

Finally, it is accepted that many such $g^*$ may exist. The fact is that, since all of them are circular, the only difference among them lies in the names of the agents. Therefore, two different Nash networks on $N$ are isomorphic. That is, if $g^*$ and $g^{*'}$ are two Nash networks on $N$, there exists a function $f:N\to N$ such that, for every $i$ and $j$,
\[
g_{i,j}^*=g_{f(i),f(j)}^{*'}.
\]
\hfill $\square$

Proposition 3 clearly indicates the close relationship between the strictness of Nash equilibria and the minimality of the number of links in the resulting structure. That is:

\textbf{Corollary.} Given a component $g$, it is a strict Nash equilibrium if and only if the number of its links is minimal.

\textit{Proof.} Suppose that $g$ is a strict Nash equilibrium, but the number of links is not minimal. Then there must be a redundant link, that is, a link that, if cut, would leave the payoff of at least one agent unchanged. Thus, there exists an agent $i$ and a deviation $g_i'$ such that
\[
\Pi_i(g_i',g_{-i})=\Pi_i(g).
\]
This is a contradiction, since we assumed that $g$ is a strict Nash equilibrium.

Since $g$ is a component, for every pair of agents $i$ and $j$,
\[
i\in N^{j;g}
\quad \text{and} \quad
j\in N^{i;g}.
\]
That is, all agents are connected. As discussed above, the minimum number of links that ensures this is $n$. The only structure with the property that all agents are connected by $n$ links is the circular network, which, according to Proposition 3, is a strict Nash equilibrium. \hfill $\square$

Even if circular networks can be identified with strict Nash equilibria, this does not mean that they are unique within the set of agents in $N$. However, they are certainly isomorphic, as shown in the following example.

\textbf{Example 6.} Let
\[
N=\{1,2,3\}
\]
with
\[
I_1=2,\quad I_2=3,\quad I_3=4.
\]
Let $g^*$ be represented by the strategy profile in Table 6.

\begin{table}[h!]
\centering
\caption{Strategy profile}
\begin{tabular}{c|ccc}
\toprule
Strategy & 1 & 2 & 3 \\
\midrule
$g_1^*$ & X & 1 & 0 \\
$g_2^*$ & 0 & X & 1 \\
$g_3^*$ & 1 & 0 & X \\
\bottomrule
\end{tabular}
\end{table}

Establish that $g^*$ is a Nash equilibrium. Consider the best response of 1 to $g_{-1}^*$. There are four options:
\[
g_1^1=(X,0,0),\quad
g_1^2=(X,1,0),\quad
g_1^3=(X,0,1),\quad
g_1^4=(X,1,1).
\]
It is found that
\[
\Pi_1(g_1^1,g_{-1}^*)=I_1=2,
\]
\[
\Pi_1(g_1^2,g_{-1}^*)=I_1+I_2+I_3-2=2+3+4-2=7,
\]
\[
\Pi_1(g_1^3,g_{-1}^*)=I_1+I_3-2=2+4-2=4,
\]
and
\[
\Pi_1(g_1^4,g_{-1}^*)=I_1+I_2+I_3-3=2+3+4-3=6.
\]
It is clear that $g_1^2$ is the best response to $g_{-1}^*$, and precisely $g_1^2=g_1^*$. A similar argument is valid for $g_2^*$ and $g_3^*$. This shows that $g^*$ is a Nash network.

On the other hand, consider the following alternative network, $g^{**}$, on $N$, represented in Table 7.

\begin{table}[h!]
\centering
\caption{Strategy profile}
\begin{tabular}{c|ccc}
\toprule
Strategy & 1 & 2 & 3 \\
\midrule
$g_1^{**}$ & X & 0 & 1 \\
$g_2^{**}$ & 1 & X & 0 \\
$g_3^{**}$ & 0 & 1 & X \\
\bottomrule
\end{tabular}
\end{table}

A quick examination shows that $g^{**}$ is also a Nash network, which for every agent in $N$ provides the same payoff:
\[
\Pi_1(g^{**})=\Pi_2(g^{**})=\Pi_3(g^{**})
=I_1+I_2+I_3-2
=9-2=7.
\]
It is easy to establish an isomorphism $f:N\to N$ between $g^*$ and $g^{**}$:
\[
f(1)=2,\quad f(2)=1,\quad f(3)=3.
\]
Then consider Table 8, obtained from the description of $g^*$ by a transposition of the rows and columns according to $f$.

\begin{table}[h!]
\centering
\caption{Relabeled strategy profile}
\begin{tabular}{c|ccc}
\toprule
Strategy & $f(1)$ & $f(2)$ & $f(3)$ \\
\midrule
$g_{f1}^{**}$ & X & 0 & 1 \\
$g_{f2}^{**}$ & 1 & X & 0 \\
$g_{f3}^{**}$ & 0 & 1 & X \\
\bottomrule
\end{tabular}
\end{table}

Notice that the structure of entries in this table is identical to that corresponding to $g^{**}$. This establishes the isomorphism between $g^*$ and $g^{**}$.

According to Lemma 1 and Proposition 3, a stable result in the strategic interaction of agents configures a circular network. We claim that it is stable because there are no incentives to cut or establish new links. Once the circular network structure has emerged, the new configuration may fail to give the same payoffs to the agents. This argument raises the question of the optimality of the result. That is, is there another configuration that can ensure better payoffs for the agents? Before answering this question negatively, two different notions of optimality must be introduced.

One represents the notion of social welfare ensured by the network. Formally, let
\[
W:G\to \mathbb{Z}
\]
be defined as
\[
W(g)=\sum_{i=1}^n \Pi_i(g)
\]
for $g\in G$. A network is said to be efficient if
\[
W(g)\geq W(g')
\]
for every $g'\in G$.

On the other hand, we have the notion of Pareto optimality. A network $g$ is said to be Pareto optimal if there is no other network $g'$ such that, for every $i\in N$,
\[
\Pi_i(g')\geq \Pi_i(g),
\]
and for at least one $i$,
\[
\Pi_i(g')>\Pi_i(g).
\]

It then follows that:

\textbf{Proposition 4.} A strict Nash network is both efficient and Pareto optimal.

\textit{Proof.} Recall that a strict Nash network $g^*$ sustains the maximum payoff for each agent:
\[
\Pi_i(g^*)=\sum_{j\in N}I_j-(n-1).
\]
Therefore,
\[
\Pi_i(g')\leq \Pi_i(g^*)
\]
for every $i\in N$ and every $g'\in G$. Thus, $g^*$ is optimal in the Pareto sense. By the same reasoning,
\[
W(g')=\sum_{i\in N}\Pi_i(g')
\leq
\sum_{i\in N}\Pi_i(g^*)=W(g^*)
\]
for every $g'\in G$. That is, $g^*$ is efficient. \hfill $\square$

\section*{Comparison with the Bala and Goyal Scheme}

As mentioned, the proposed model shares several characteristics with that of Bala and Goyal (2000), hereafter BG. But, as we shall see, the intuition is very different in one case and in the other. Moreover, the results that follow, even if there is some similarity among them, are reached on the basis of different concepts.

To organize the discussion, we introduce the notion of payoffs used in BG. Consider two definitions already given:
\[
N^{g_i}=\{k\in N\mid g_{i,k}=1\}
\]
is the set of agents with whom $i$ establishes a direct link according to her strategy $g_i$, while the set of agents accessed, directly or otherwise, by $i$ is
\[
N^{i;g}=\{k\in N\mid k\to i\}\cup\{i\}.
\]
On these two sets, BG define two functions:
\[
\delta_i^d=|N^{g_i}|
\]
and
\[
\delta_i(g)=|N^{i;g}|,
\]
which indicate, respectively, the number of agents to whom $i$ has a direct link and the number of agents to whom $i$ is connected, directly or indirectly. In BG's original presentation, $\delta_i^d$ is denoted as $\mu_i^d$, while $\delta_i$ is $\mu_i$. Here they have been reformulated to avoid confusion.

BG consider the following payoff function:
\begin{equation}
\Pi_i^{BG}(g)=\delta_i(g)-\delta_i^d(g)c,
\tag{9}
\end{equation}
where $c$ is the cost of establishing each link. That is, the payoffs of $i$ are the number of agents whose information can be accessed by her, minus the cost of the direct links established according to $g$.

\textbf{Example 2, second reformulation.} Suppose again that the information held by the agents is:
\[
I_1=2,\quad I_2=2,\quad I_3=4,\quad I_4=3,\quad I_5=3.
\]
Then, under the strategy $g$, we have in BG, assuming $c=1$:
\[
\Pi_1^{BG}(g)=\delta_1(g)-\delta_1^d(g)=5-2=3,
\]
\[
\Pi_2^{BG}(g)=\delta_2(g)-\delta_2^d(g)=4-1=3,
\]
\[
\Pi_3^{BG}(g)=\delta_3(g)-\delta_3^d(g)=3-2=1,
\]
\[
\Pi_4^{BG}(g)=\delta_4(g)-\delta_4^d(g)=1-0=1,
\]
\[
\Pi_5^{BG}(g)=\delta_5(g)-\delta_5^d(g)=1-0=1.
\]

It can be noted here that, for example, if $g_{1,5}=0$, agent 1 could improve her benefit, obtaining 10 instead of 9, because she could continue to have access to the information of agent 5 but using one less link. The same remains true in the BG case, which in this particular case would increase her benefit from 3 to 4.

However, if, for example, agent 3 does not contact agent 5, agent 1 would improve her payoff under $\Pi_1$ from 9 to 10, whereas in the case of $\Pi_1^{BG}$ her payoff remains equal to 3.

Moreover, note that while
\[
\Pi_3^{BG}(g)=1=\Pi_4^{BG}(g),
\]
we have
\[
\Pi_3(g)=8>3=\Pi_4(g).
\]

The differences in the payoffs shown in this example clearly display the different intuitions behind $\Pi_i(g)$ and $\Pi_i^{BG}(g)$. While the former depends on the value of the information available to the agents, BG base it on the number of agents accessed. The costs are also different. BG only consider the costs of establishing direct links, whereas in our case the cost of a path is shared by all agents that make it up.

To analyze the existence of equilibria, BG generalize $\Pi_i^{BG}$ by means of the function
\begin{equation}
\Phi(\delta_i(g),\delta_i^d(g)),
\tag{10}
\end{equation}
increasing in the first argument and decreasing in the second. With this function, BG prove the following statements:

\begin{itemize}
\item A Nash network is either empty or minimally connected. Proposition 3.1, p. 1194.
\item A strict Nash network is either empty or circular. Proposition 3.2, p. 1195.
\end{itemize}

Moreover:

\begin{enumerate}
\item If
\[
\Phi(m+1,m)>\Phi(1,0)
\]
for some
\[
m\in\{1,\ldots,n-1\},
\]
the unique strict Nash network is the circular network.

\item If
\[
\Phi(m+1,m)\leq \Phi(1,0)
\]
for all $m$, and
\[
\Phi(n,1)>\Phi(1,0),
\]
both the circular network and the empty network are strict Nash networks.

\item If
\[
\Phi(m+1,m)\leq \Phi(1,0)
\]
for all $m$, and
\[
\Phi(n,1)\leq \Phi(1,0),
\]
the empty network is the unique strict Nash network.
\end{enumerate}

This last result can be evaluated when
\[
\Phi(\delta_i(g),\delta_i^d(g))=\delta_i(g)-\delta_i^d(g)c.
\]
Thus, case 1 reduces to
\[
m+1-mc>1,
\]
that is,
\[
1-c>0,
\]
or
\[
c<1.
\]
Case 2 indicates that while
\[
c\geq 1,
\]
it also holds that
\[
n-c>1,
\]
that is,
\[
c<n-1.
\]
Finally, case 3 occurs when
\[
c\geq n-1.
\]
Thus, these results indicate that the only strict Nash networks are circular when $c<1$, whereas empty networks are the only Nash networks when $c>n-1$. For
\[
c\in[1,n-1],
\]
both the circular network and the empty network are strict Nash equilibria.

To compare the results with those of BG, it is necessary to identify the respective payoff functions, that is:
\begin{equation}
\Pi_i(g)=
\sum_{j\in N^{i;g}}I_j-\mu_i(g)
=
\delta_i(g)-\delta_i^d(g)c
=
\Pi_i^{BG}(g).
\tag{11}
\end{equation}

The simplest way in which this can occur is if
\[
I_j=1
\]
for each
\[
j\in N^{i;g},
\]
and
\[
\mu_i(g)=\delta_i^d(g)c.
\]
In particular, for circular networks it is the case that, for each $i$,
\[
N^{i;g}=N,\quad \mu_i(g)=n-1,\quad \delta_i^d(g)=1.
\]
To keep both payoff functions equal, the cost of each link must be
\[
c=n-1.
\]

But then, although the payoffs are the same, our approach identifies the circular network as the unique strict Nash network. In BG, by contrast, the unique strict Nash network is the empty network. In other words, the differences between the respective payoff functions lead to differences in equilibria.

\section*{A Final Discussion}

This paper has presented a model of network formation in the form of a noncooperative game, where agents decide to whom to link by comparing the net benefits of their actions. Decisions are made simultaneously, and therefore no dynamic scheme is required, as appears in Bala and Goyal (2000). In any case, in a model in which agents are not myopic, that is, as in our scheme, and face higher costs of establishing initial links, dynamic processes that converge to the circular network can also be proposed; see Watts (2000).

In this analytical framework, heterogeneity only means that each agent is endowed with some particular information that is valuable to other agents. This assumption leads to an increase in the benefit of entering the network, making participation in the network always more valuable than isolation. On the other hand, costs tend to be higher in our scheme because it is assumed that agents contribute to paying the costs of the links in the paths that carry information to them.

An interesting result is that circular networks emerge here as Nash equilibria that sustain structures with a minimum number of links. This supports the intuition that strict Nash networks and the existence of a minimum number of connections are equivalent properties.

\section*{References}

Bala, V., and Goyal, S. ``A Noncooperative Model of Network Formation.'' \textit{Econometrica}, 68, 2000, pp. 1181--1229.  \url{https://doi.org/10.1111/1468-0262.00155}

Dutta, B., and Jackson, M. ``The Stability and Efficiency of Directed Communication Networks.'' \textit{Review of Economic Design}, 5, 2000, pp. 251--272. \url{https://doi.org/10.1007/PL00013688}

Dutta, B., and Jackson, M. \textit{On the Formation of Networks and Groups}. In B. Dutta and M. Jackson, eds., \textit{Models of Strategic Formation of Networks and Groups}. Springer-Verlag, New York, 2001.

Dutta, B., van den Nouweland, A., and Tijs, S. ``Link Formation in Cooperative Situations.'' \textit{International Journal of Game Theory}, 27, 1998, pp. 245--255. \url{https://doi.org/10.1007/s001820050070}

Dutta, B., and Mutuswami, S. ``Stable Networks.'' \textit{Journal of Economic Theory}, 76, 1997, pp. 322--344. DOI: 10.1006/jeth.1997.2306

Jackson, M., and Wolinsky, A. ``A Strategic Model of Social and Economic Networks.'' \textit{Journal of Economic Theory}, 71, 1996, pp. 44--74. DOI: 10.1006/jeth.1996.0108

Qin, C. Z. ``Endogenous Formation of Cooperative Structures.'' \textit{Journal of Economic Theory}, 69, 1996, pp. 218--226. DOI: 10.1006/jeth.1996.0047

Slikker, M., and van den Nouweland, A. ``A One-Stage Model of Link Formation and Payoff Division.'' \textit{Games and Economic Behavior}, 34, 2001, pp. 153--175. DOI: 10.1006/game.1999.0785

Tanenbaum, A. \textit{Computer Networks}. Prentice Hall, Englewood Cliffs, New Jersey, 1989.

Wasserman, S., and Faust, K. \textit{Social Network Analysis}. Cambridge University Press, New York, 1994.

Watts, A. ``Non-myopic Formation of Circle Networks.'' \textit{Economic Letters}, 74, 2002, pp. 277--282. \url{https://doi.org/10.1016/S0165-1765(01)00540-7}

\end{document}